\documentclass[aps,preprint]{revtex4}%
\usepackage{amsfonts}
\usepackage{amsmath}
\usepackage{amssymb}
\usepackage{graphicx}
\usepackage{xcolor}%
\providecommand{\U}[1]{\protect\rule{.1in}{.1in}}

\newcommand{\beq}{\begin{equation}}
\newcommand{\eeq}{\end{equation}}
\newcommand{\bea}{\begin{eqnarray}}
\newcommand{\eea}{\end{eqnarray}}
\begin{document}
\preprint{ }
\title{Finite Temperature Scaling in Density Functional Theory}
\author{James W.\ Dufty}
\affiliation{Department of Physics, University of Florida, Gainesville FL 32611}
\author{S.B.\ Trickey}
\affiliation{Quantum Theory Project, Dept.\ of Physics and Dept.\ of Chemistry, University
of Florida, Gainesville FL 32611}

\begin{abstract}
A previous analysis of scaling, bounds, and inequalities for the
non-interacting functionals of thermal density functional theory is extended
to the full interacting functionals. The results are obtained from analysis of
the related functionals from the equilibrium statistical mechanics of
thermodynamics for an inhomogeneous system. Their extension to the functionals
of density functional theory is described.

\end{abstract}
\date{(revised November 5, 2015)}
\maketitle

\section{Commendation}

\label{sec0} Andreas Savin has contributed several innovative analyses and
insights to the formal and practical development of density functional theory
(DFT). They are marked with a characteristic intellectual style which often
illuminates points which few, if any, others had considered. One of us (SBT)
has had many conversations of that nature with Andreas - and typically came
away from each having learned much. In the context of the present
contribution, we note one particular Savin paper: \textquotedblleft On
Degeneracy, Near-degeneracy, and Density Functional Theory\textquotedblright%
\cite{Savin96}. It explores ensemble and many-determinant forms of DFT in the
context of dissociation limits, symmetries, and invariants. Ensembles, of
course, are central to statistical mechanics, hence the question of their
implications for the behavior of finite-temperature functionals automatically
arises. Here we analyze the scaling behavior of finite-T ensembles. Though the
issues addressed here differ from those in Ref.\ \cite{Savin96}, the
underlying approach is closely related, namely the exploration of invariance
properties. We hope that Andreas as well as others find value in it and salute
him on his formal (but, we trust, not actual) retirement.

\section{Introduction}

\label{sec1} Thermal DFT is a formally exact structure for the prediction of
thermodynamic properties of a quantum or classical system
\cite{Mermin65,YangParrBook,DreizlerGrossBook,KryachkoLudena,Eschrig,LudenaKarasiev2002,FilohaisEtAl03,EngelDreizler11,Evans92,Lowen94,Dharma95,Lutsko10}%
. Its application requires specification of certain functionals of the density
which are not given \emph{a priori} and for which no mechanical recipe
(\textit{e.g.}\ perturbation expansion) is provided by the existence theorems.
Development of accurate approximate functionals therefore remains the primary
challenge for accurate implementation of DFT for a given problem class. An
example of current interest is the subject of warm dense matter (WDM). The
state conditions of WDM include densities and temperatures for which both
traditional zero-temperature solid state or molecular forms and
high-temperature plasma forms fail \cite{WDM,IPAMbook}. In this context it is
important to have exact properties of the free-energy density functionals for
guidance in construction of approximations. One category of exact results that
has proved fruitful at zero temperature is scaling laws, that is,
relationships for how the functionals change when the density is transformed
under a uniform coordinate scaling which preserves total particle number
\cite{LevyPerdew85}. Recently, we addressed this problem for the functionals
of a non-interacting system at finite temperature \cite{DuftyTrickey11}. Here,
those results are extended to the corresponding functionals for the
interacting system and, as a consequence, for the exchange- correlation components.

Similar results have been obtained by Pittalis \textit{et al.}
\cite{Pittalis11}. The primary simplifying feature of the approach used here
and in Ref.\ \onlinecite{DuftyTrickey11} is recognition of the origin of the
density functionals in the statistical mechanics of a non-uniform system at
equilibrium \cite{Thermo}. The scaling properties then follow directly from
their explicit representations as equilibrium ensemble averages via invariance
of corresponding dimensionless forms. One consequence is that coupling
constant scaling (\textit{i.e.}, electronic charge scaling) appears
intrinsically in the dimensional analysis, rather than as a separate
consequence of adiabatic connection as in Ref.\ \onlinecite{Pittalis11}.

In the next section, the thermodynamic context is noted and the density
functionals for free energy, internal energy, and entropy are defined. In
section \ref{sec3} the treatment is specialized to the important case of
systems of electrons and ions. A coordinate scaling transformation for the
electron number density is extended to include a scaling of the temperature
and the Coulomb coupling constant that leaves the dimensionless functionals
invariant. The consequences of this invariance are the scaling laws of
interest. These results are discussed further in section \ref{sec4}.

\section{Equilibrium statistical mechanics and density functional theory}

\label{sec2}

\subsection{Thermodynamics}

Here we recall the relevant functionals of local thermodynamics on the basis
of their statistical mechanical definitions in the grand canonical ensemble.
Their relationship to the functionals and variational context of density
functional theory then is noted. Recently, Eschrig \cite{Eschrig10} discussed
this relationship for the general quantum case and a pedagogical version for
the classical case is in Ref.\ \onlinecite{Caillol02}. The relationship
provides the basis for establishing scaling laws from dimensional analysis and
their relationship to thermodynamic transformations.

The grand canonical ensemble describes a system at equilibrium, exchanging
energy and particles with its surroundings. The thermodynamic parameters are
the temperature ${\mathrm{T}}=1/k_{B}\beta,$ volume $V$, and local chemical
potential $\mu(\mathbf{r})=\mu-v_{ext}(\mathbf{r})$. The presence of an
external potential $v_{ext}(\mathbf{r})$ implies that, in general, the system
is inhomogeneous (lacks translational invariance). The thermodynamic
properties are defined in terms of the grand potential
\begin{equation}
\beta\Omega(\beta\mid\mu)\equiv-\ln\sum_{N=0}^{\infty}Tr^{(N)}e^{-\beta\left(
\widehat{H}-\int d\mathbf{r}\mu(\mathbf{r})\widehat{n}(\mathbf{r})\right)
}\,, \label{2.1}%
\end{equation}
where $\widehat{H}$ is the Hamiltonian operator (see below; note the usual
assumption that $\widehat{H}$ is bounded below), $\widehat{n}(\mathbf{r})$ is
the particle number density operator%
\begin{equation}
\widehat{n}(\mathbf{r})=\sum\limits_{i=1}^{N}\delta\left(  \mathbf{r-}%
\widehat{\mathbf{r}}_{i}\right)  , \label{2.2}%
\end{equation}
and $\widehat{\mathbf{r}}_{i}$ is the position operator for the $i^{th}$
particle. The traces are taken over $N$-particle Hilbert space with an
appropriate restriction on the symmetry of states (Bosons or Fermions). The
grand potential and pressure $p(\beta\mid\mu)$ are proportional: $p(\beta
\mid\mu)V=-\Omega(\beta\mid\mu)$.

The primitive functional of interest in the present context is the grand
potential $\beta\Omega(\beta\mid\cdot)$ defined over the class of functions
$\mu(\mathbf{r})$ for which the right side of (\ref{2.1}) exists. An important
property that follows from this definition is its concavity%
\begin{equation}
\beta\Omega(\beta\mid\alpha\mu_{1}+\left(  1-\alpha\right)  \mu_{2}%
)>\alpha\beta\Omega(\beta\mid\mu_{1})+\left(  1-\alpha\right)  \beta
\Omega(\beta\mid\mu_{2}) \label{2.3}%
\end{equation}
for $0<\alpha<1$ and arbitrary $\mu_{1}(\mathbf{r})$ and $\mu_{2}(\mathbf{r})$
in the defining class. A closely related functional is the local number
density defined by
\begin{equation}
n(\mathbf{r,}\beta\mid\mu)\equiv-\frac{\delta\Omega(\beta\mid\mu)}{\delta
\mu\left(  \mathbf{r}\right)  }. \label{2.4}%
\end{equation}
Whenever this last definition is invertible it identifies the chemical
potential as a functional of the density%
\begin{equation}
\mu\left(  \mathbf{r}\right)  \equiv\mu(\mathbf{r,}\beta\mid n). \label{2.5}%
\end{equation}
Thus, at thermodynamical equilibrium there are dependent pairs $\mu$, $n$ such
that one determines the other. It then is possible to define a change of
variables from $\beta,V,\mu\left(  \mathbf{r}\right)  $ to $\beta,V,n\left(
\mathbf{r}\right)  $. This is effected by the Legendre transformation
\begin{align}
F(\beta &  \mid n)\equiv\left[  \Omega(\beta\mid\mu)-\int d\mathbf{r}%
\mu\left(  \mathbf{r}\right)  \frac{\delta\Omega(\beta\mid\mu)}{\delta
\mu\left(  \mathbf{r}\right)  }\right]  _{\mu(\mathbf{r,}\beta\mid
n)}\nonumber\\
&  =\left[  \Omega(\beta\mid\mu)+\int d\mathbf{r}\mu\left(  \mathbf{r}\right)
n(\mathbf{r})\right]  _{\mu(\mathbf{r,}\beta\mid n)} \label{2.6}%
\end{align}
whose differential with respect to $\mu\left(  \mathbf{r}\right)  $ at
constant $n\left(  \mathbf{r}\right)  $ manifestly vanishes. Hence,
$F(\beta\mid\cdot)$ is a universal functional of the density, in the sense
that it is independent of $\mu\left(  \mathbf{r}\right)  $ (hence of $v_{ext}%
$). Also $F(\beta\mid n)$, the thermodynamic free energy, is a convex
functional of $n\left(  \mathbf{r}\right)  $.
\begin{equation}
\beta F(\beta\mid\alpha n_{1}+\left(  1-\alpha\right)  n_{2})<\alpha\beta
F(\beta\mid n_{1})+\left(  1-\alpha\right)  \beta F(\beta\mid n_{2})
\label{2.7}%
\end{equation}
for $0<\alpha<1$ and arbitrary $n_{1}(\mathbf{r})$ and $n_{2}(\mathbf{r}).$

The equivalent relationship between the pair $\mu$, $n$ of (\ref{2.6}) now is
expressed as%
\begin{equation}
\mu(\mathbf{r})=\frac{\delta F(\beta\mid n)}{\delta n\left(  \mathbf{r}%
\right)  } \; . \label{2.7a}%
\end{equation}
Other thermodynamic functionals of interest are defined in terms of the
foregoing two. For example, the energy and entropy functionals are%
\begin{equation}
U(\beta\mid n)=\frac{\partial\beta\Omega(\beta\mid\mu)}{\partial\beta}%
\mid_{\beta\mu},\mathbf{\hspace{0.2in}}TS(\beta\mid n)=U(\beta\mid
n)-F(\beta\mid n) \; . \label{2.8}%
\end{equation}

\subsection{Connection with density functional theory}

The connection to DFT is established by defining a bi-linear functional
${\Omega}(\beta\mid\mu,n)$ which is closely related to $\Omega(\beta\mid\mu)$.
The new functional is defined for the density $n(\mathbf{r})$ and for an
\textit{independently specified} $\mu(\mathbf{r})$,
\begin{equation}
\Omega(\beta\mid\mu,n)\equiv F(\beta\mid n)-\int d\mathbf{r}\mu\left(
\mathbf{r}\right)  n(\mathbf{r})\;. \label{2.9}%
\end{equation}
The density functional $F(\beta\mid\cdot)$ in this definition is the same as
the thermodynamic functional given by (\ref{2.6}). However, $\Omega(\beta
\mid\mu,n)$ differs from $\Omega(\beta\mid\mu)$ because
\begin{equation}
\int d\mathbf{r}\mu\left(  \mathbf{r}\right)  n(\mathbf{r})\neq-\int
d\mathbf{r}\mu\left(  \mathbf{r}\right)  \frac{\delta\Omega(\beta\mid\mu
)}{\delta\mu\left(  \mathbf{r}\right)  } \label{2.10}%
\end{equation}
for \textit{separately and arbitrarily specified} $\mu\left(  \mathbf{r}%
\right)  $ and $n\left(  \mathbf{r}\right)  $. The same statement is true if
$\Omega(\beta\mid\mu, n)$ appears on the RHS of (\ref{2.10}). The two
functionals, $\Omega(\beta\mid\mu,n)$ and $\Omega(\beta\mid\mu)$, do become
equal when $\mu\left(  \mathbf{r}\right)  $, $n\left(  \mathbf{r}\right)  $
are the matched pair related by Eq.\ (\ref{2.7a}). Furthermore, the special
density which provides that pair for given $\mu\left(  \mathbf{r}\right)  $ is
that density $\overline{n}$ which minimizes
\begin{equation}
\min_{n}\Omega(\beta\mid n,\mu)\equiv\Omega(\beta\mid\mu,\overline{n}%
)=\Omega(\beta\mid\mu). \label{2.11}%
\end{equation}
The Euler equation associated with (\ref{2.11}) which defines the special
density is
\begin{equation}
\mu(\mathbf{r})=\frac{\delta F(\beta\mid n)}{\delta n\left(  \mathbf{r}%
\right)  }\Big\vert_{n=\overline{n}}. \label{2.12}%
\end{equation}
This recovers the thermodynamic relationship (\ref{2.7a}) as expected.
Equation (\ref{2.9}) defines the fundamental functional of DFT, while
(\ref{2.11}) and (\ref{2.12}) are the variational applications of it
\cite{Mermin65}. Those follow from the convexity properties of the
thermodynamic functionals. In particular, it can be shown that $\Omega
(\beta\mid\mu,n)$ is a convex functional of $n$ for which the minimum is the
desired special density \cite{Ruelle1969}.

The customary formulation and application of DFT is to determine the density
for given $\mu\left(  \mathbf{r}\right)  $. From the foregoing thermodynamic
discussion it is evident, however, that $n\left(  \mathbf{r}\right)  $ and
$\mu\left(  \mathbf{r}\right)  $ have dual roles at equilibrium with
corresponding thermodynamic potentials $F(\beta\mid n)$ and $\Omega(\beta
\mid\mu)$, respectively. It might be expected, for example as a parallel to
the original Hohenberg-Kohn \cite{H_Kohn} bijectivity proof at T=0K
\cite{YangParrBook,DreizlerGrossBook}, that a complementary version of thermal
DFT could be formulated in terms of variation of $\mu\left(  \mathbf{r}%
\right)  $ at given density $n\left(  \mathbf{r}\right)  $. This is indeed the
case \cite{Liebref,Caillol02,Eschrig10}. Define the bi-linear functional for
given $n\left(  \mathbf{r}\right)  $ but arbitrary $\mu\left(  \mathbf{r}%
\right)  $
\begin{equation}
F(\beta\mid n,\mu)\equiv\Omega(\beta\mid\mu)+\int d\mathbf{r}\mu\left(
\mathbf{r}\right)  n(\mathbf{r})\;.\label{2.13}%
\end{equation}
The functional $\Omega(\beta\mid\cdot)$ is the same as the grand potential
functional given by (\ref{2.1}). However, $F(\beta\mid n,\mu)$ differs from
$F(\beta\mid n)$ because of the inequality (\ref{2.10}) for general
$\mu\left(  \mathbf{r}\right)  $. It can be shown that $F(\beta\mid n,\mu)$ is
a concave functional of $\mu$, and%
\begin{equation}
\sup_{\mu}F(\beta\mid n,\mu)=F(\beta\mid n,\overline{\mu})=F(\beta\mid
n),\label{2.14}%
\end{equation}
where the extremum is attained for $\mu\left(  \mathbf{r}\right)
=\overline{\mu}\left(  \mathbf{r}\right)  $ determined from
\begin{equation}
n(\mathbf{r})=-\frac{\delta\Omega(\beta\mid\mu)}{\delta\mu\left(
\mathbf{r}\right)  }\Big\vert_{\mu=\overline{\mu}}.\label{2.15}%
\end{equation}
This is the required thermodynamic relationship given in (\ref{2.4}), while
the last equality of (\ref{2.14}) shows that the thermodynamic free energy is
recovered for this particular value of the chemical potential. Equations
(\ref{2.13}) - (\ref{2.15}) comprise a representation of density functional
theory that is fully equivalent to the customary form displayed in
Eqs.\ (\ref{2.9}), (\ref{2.11}), and (\ref{2.12}). For the T= 0K analogue, see
Lieb \cite{Liebref}.

In summary, this section has defined the relevant functionals for the
thermodynamics of an inhomogeneous system within the framework of the grand
ensemble of equilibrium statistical mechanics. The simple relationship of the
central ingredients of density functional theory to those equilibrium
functionals then was indicated. The advantage of this perspective is that the
functionals have an unambiguous representation within statistical mechanics,
for both formal and practical analysis.

\section{Scaling properties of functionals}

\label{sec3}

Most commonly, the systems of interest for DFT applications are comprised of
electrons and ions, denoted in the following by subscript ``\textit{e}'' and
``\textit{i}''. For simplicity without loss of generality, we restrict
discussion to a single ion species. The grand potential reads
\begin{equation}
\beta\Omega(\beta,\mu_{e},\mu_{i})\equiv-\ln\sum_{N_{e}=0,N_{i}=0}^{\infty
}Tr^{(N_{e})}Tr^{(N_{i})}e^{-\beta\left(  \widehat{H}_{e}+\widehat{H}%
_{i}+\widehat{H}_{ei}-\mu_{e}N_{e}-\mu_{i}N_{i}\right)  }\, . \label{3.1}%
\end{equation}
The two chemical potentials are related by charge neutrality%
\begin{equation}
q_{e}N_{e}(\beta,\mu_{e},\mu_{i})+q_{i}N_{i}(\beta,\mu_{e},\mu_{i})=0,
\label{3.2}%
\end{equation}
or%
\begin{equation}
q_{e}\frac{\partial\Omega(\beta,\mu_{e},\mu_{i})}{\partial\mu_{e}}+q_{i}%
\frac{\partial\Omega(\beta,\mu_{e},\mu_{i})}{\partial\mu_{i}}=0. \label{3.3}%
\end{equation}
In the following discussion, it is assumed that $\mu_{e}$ is given and
$\mu_{i}$ is determined from this charge neutrality condition.

Because of the comparatively large mass of the ions, typically the ionic
thermal de Broglie wavelength $\lambda_{i}$ and average inter-ionic separation
$r_{0}$ are much different: $\lambda_{i}<<r_{0}$. The ions therefore can be
treated in a semi-classical limit. The motive for doing so is to have a purely
electronic DFT with the ions providing an external potential. The average over
the electron subsystem can be performed formally to give
\begin{equation}
\beta\Omega(\beta,\mu_{e},\mu_{i})\rightarrow-\ln\sum_{N_{i}=0}^{\infty}%
\frac{1}{\lambda_{i}^{3N_{i}}N_{i}!}\int d\mathbf{X}_{i}e^{-\beta\left(
U_{ii}+\Omega_{e}(\beta,q_{e}\mid\mu)-\mu_{i}N_{i}\right)  }. \label{3.4}%
\end{equation}
Here $U_{ii}$ is the ion-ion Coulomb repulsion energy and the set of ionic
coordinates $\left\lbrace {\mathbf{X}}_{i,1} \ldots{\mathbf{X}}_{i,N_{i}%
}\right\rbrace $ is denoted ${\mathbf{X}}_{i}$. The electronic grand potential
$\Omega_{e}$ that appears in (\ref{3.4}) is defined by
\begin{equation}
\beta\Omega_{e}(\beta,q_{e}\mid\mu)=-\ln\sum_{N_{e}=0}^{\infty}Tr^{(N_{e}%
)}e^{-\beta\left(  \widehat{H}_{e}-\int d\mathbf{r}\mu(\mathbf{r})\widehat
{n}_{e}(\mathbf{r})\right)  }, \label{3.5}%
\end{equation}
where the dependence on the electronic charge $q_{e}$ now has been made
explicit for purposes of the scaling analysis of the next subsection. The
electronic local chemical potential
\begin{equation}
\mu(\mathbf{r})=\mu_{e}-v_{ext}(\mathbf{r}),\hspace{0.2in}v_{ext}%
(\mathbf{r})=\sum_{\eta=1}^{N_{i}}\frac{q_{e}q_{i}}{|\mathbf{r}-\mathbf{X}%
_{i\eta}|}. \label{3.6}%
\end{equation}
The dependence of $\beta\Omega_{e}$ and $\mu$ on the ion coordinates
$\mathbf{X}_{i}$ has been left implicit in this notation.

In this approximation, the calculation of the thermodynamic properties for the
electron - ion system is separated into two parts. First the quantum
mechanical treatment of the electrons is addressed through the calculation of
$\beta\Omega_{e}(\beta\mid\mu_{e})$, the task of DFT. Next, the classical ion
average is performed (typically in the Born-Oppenheimer approximation). That
can be done via molecular dynamics simulation under the assumption of
ergodicity. The effective ion forces for such a simulation are%
\begin{align}
{\mathrm{F}}_{i}  &  =-\nabla_{X_{i}}\left(  U_{ii}+\Omega_{e}(\beta,q_{e}%
\mid\mu)\right) \nonumber\\
&  =-\nabla_{\mathbf{X}_{i}}U_{ii}-\int d\mathbf{r}n_{e}(\mathbf{r,}\beta
\mid\mu)\nabla_{\mathbf{X}_{i}}v_{ext}(\mathbf{r}), \label{3.7}%
\end{align}
where the definition of the density, (\ref{2.4}), has been used in the second
equality. It can be shown that this same result is obtained from the
microscopic ion force averaged over the electron subsystem. (As an aside, we
remark that such a simulation calculates one member of the grand ensemble. For
large systems, the fluctuations about the average number of electrons are
small, hence so are the fluctuations around the average number of ions. Thus a
single simulation can be sufficient and operationally equivalent to working in
the canonical ensemble. More generally, fluctuation corrections can be
calculated. These issues of precise relationship between theory and
simulational practice are worth keeping in mind but are not of consequence
here. See Ref.\ \cite{ChakrabortyEtAl2015} for an analysis.)

The remainder of this work addresses only the electron subsystem, namely the
calculation of $\Omega_{e}(\beta,q_{e}\mid\mu)$ from DFT. Therefore, from now
on we drop the subscript \textquotedblleft$e$\textquotedblright\ on the grand
potential and charge and write $\Omega(\beta,q\mid\mu)$ and do the same with
the associated free energy functional $F(\beta,q\mid n)$.

\subsection{Scaling properties of $\Omega(\beta,q\mid\mu)$ and $F(\beta,q\mid
n)$}

Begin, as before, with the thermodynamic functionals and introduce the
dimensionless variables/operators
\begin{equation}
\nu(\mathbf{r}^{\ast})=\beta\mu(\mathbf{r}),\hspace{0.2in}\widehat{n}^{\ast
}(\mathbf{r}^{\ast})=\lambda^{3}\widehat{n}(\mathbf{r}),\hspace{0.2in}%
\widehat{\mathbf{y}}_{\alpha}\equiv\sqrt{\frac{\beta}{2m}}\widehat{\mathbf{p}%
}_{\alpha},\hspace{0.2in}\widehat{\mathbf{r}}_{\alpha}^{\ast}=\frac
{\widehat{\mathbf{r}}_{\alpha}}{\lambda}\,,\hspace{0.2in}\xi=\beta
q^{2}/\lambda\label{3.8}%
\end{equation}
with $\widehat{\mathbf{p}}_{\alpha}$, $\widehat{\mathbf{r}}_{\alpha}$ the
momentum and position operators respectively for electron $\alpha$ and
$\lambda$ the electron thermal de Broglie wavelength%
\begin{equation}
\lambda\left(  \beta\right)  =\left(  \frac{2\pi\beta\hbar^{2}}{m}\right)
^{1/2}\,. \label{3.9}%
\end{equation}
Conversion to these dimensionless variables gives the dimensionless grand
potential
\begin{equation}
\Omega^{\ast}(\xi\mid\nu)=\beta\Omega(\beta,q\mid\mu)=-\ln\sum_{N=0}^{\infty
}Tr^{(N)}e^{-\left(  \widehat{H}^{\ast}-\int d\mathbf{r}^{\ast}\nu
(\mathbf{r}^{\ast})\widehat{n}^{\ast}(\mathbf{r}^{\ast})\right)  }\,.
\label{3.10}%
\end{equation}
The dimensionless Hamiltonian is%
\begin{equation}
\widehat{H}^{\ast}=\beta\widehat{H}=\sum_{\alpha=1}^{N}\widehat{y}_{\alpha
}^{2}+\tfrac{1}{2}\xi\sum_{\alpha\neq\eta=1}^{N}\frac{1}{|\hat{\mathbf{r}%
}_{\alpha}^{\ast}-\hat{\mathbf{r}}_{\eta}^{\ast}|}. \label{3.11}%
\end{equation}
The dimensionless electron number density follows from
\begin{equation}
n^{\ast}(\mathbf{r}^{\ast}\mathbf{,}\xi\mid\nu)=-\frac{\delta\Omega^{\ast}%
(\xi\mid\nu)}{\delta\nu\left(  \mathbf{r}^{\ast}\right)  }\Big\vert_{\xi}\,.
\label{3.12}%
\end{equation}
Assuming invertibility as usual, this relationship establishes $\nu
(\mathbf{r}^{\ast})$ as a functional of $n^{\ast}(\mathbf{r}^{\ast})$
\begin{equation}
\nu(\mathbf{r}^{\ast})=\nu(\mathbf{r}^{\ast},\xi\mid n^{\ast}). \label{3.13}%
\end{equation}

The dimensionless free energy functional of the density is defined by the
Legendre transformation
\begin{equation}
F^{\ast}(\xi\mid n^{\ast})=\beta F(\beta,q\mid n)=\Omega^{\ast}(\xi\mid
\nu)+\int d\mathbf{r}^{\ast}\nu(\mathbf{r}^{\ast},\xi\mid n^{\ast})n^{\ast
}(\mathbf{r}^{\ast})\;. \label{3.14}%
\end{equation}
The corresponding dimensionless internal energy and entropy functionals are%
\begin{equation}
U^{\ast}(\xi\mid n^{\ast})=\beta U(\beta,q\mid n)=\beta\frac{\partial
\Omega^{\ast}(\beta,q\mid\mu)}{\partial\beta}\mid_{\nu} \label{3.15}%
\end{equation}
\ \
\begin{equation}
S^{\ast}(\xi\mid n^{\ast})=k_{B}^{-1}S(\beta,q\mid n)=U^{\ast}(\xi\mid
n^{\ast})-F^{\ast}(\xi\mid n^{\ast})\;. \label{3.16}%
\end{equation}

Now consider the usual uniform coordinate scaling transformation of the
density and volume,
\begin{equation}
n(\mathbf{r})\rightarrow n_{\gamma}(\mathbf{r})=\gamma^{3}n(\gamma
\mathbf{r}),\hspace{0.2in}V\rightarrow\gamma^{-3}V \;, \label{3.17}%
\end{equation}
with $\gamma$ a real, positive constant. Include scalings of both $\beta$ and
$q$ in the transformation as well:
\begin{equation}
\beta\rightarrow\gamma^{-2}\beta,\;\hspace{0.2in}q\rightarrow\gamma^{1/2}q \;.
\label{3.19a}%
\end{equation}
The result is to leave $\xi$ invariant%
\begin{equation}
\xi=\beta q^{2}/\lambda\rightarrow\gamma^{-2}\beta\left(  \gamma
^{1/2}q\right)  ^{2}/\lambda\left(  \gamma^{-2}\beta\;\right)  =\xi\;,
\label{3.19c}%
\end{equation}
and preserve the total number of particles
\begin{equation}
\int_{V}n(\mathbf{r})d\mathbf{r}=N=\int_{\gamma^{-3}V}n_{\gamma}%
(\mathbf{r})d\mathbf{r}\,. \label{3.18}%
\end{equation}
Also $n^{\ast}(\mathbf{r}^{\ast})$ is invariant%
\begin{align}
n^{\ast}(\mathbf{r}^{\ast})  &  \equiv\lambda^{3}\left(  \beta\right)
n(\lambda\left(  \beta\right)  \mathbf{r}^{\ast})\rightarrow\lambda^{3}\left(
\gamma^{-2}\beta\right)  \gamma^{3}n(\gamma\mathbf{r})=\lambda\left(
\gamma^{-2}\beta\right)  ^{3}\gamma^{3}n(\gamma\lambda\left(  \gamma^{-2}%
\beta\right)  \mathbf{r}^{\ast\ast})\,\nonumber\\
&  =\lambda^{3}\left(  \beta\right)  n(\lambda\left(  \beta\right)
\mathbf{r}^{\ast\ast})=n^{\ast}(\mathbf{r}^{\ast\ast}) \label{4.9}%
\end{align}
where $\mathbf{r}^{\ast}\rightarrow\mathbf{r}^{\ast\ast}=\mathbf{r}%
/\lambda\left(  \gamma^{-2}\beta\right)  $ is the dimensionless coordinate in
the transformed system.

These invariances for $\xi$ and $n^{\ast}$ then imply the invariance of the
dimensionless free energy $F^{\ast}(\xi\mid n^{\ast})$ which is a function or
functional of only these properties. Consequently the scaling law for the free
energy functional is obtained
\begin{equation}
\beta F(\beta,q\mid n)=\gamma^{-2}\beta F(\gamma^{-2}\beta,\gamma^{1/2}q\mid
n_{\gamma}). \label{3.22}%
\end{equation}
It also follows that $\nu(\mathbf{r}^{\ast},\xi\mid n^{\ast})\rightarrow
\nu(\mathbf{r}^{\ast\ast},\xi\mid n^{\ast})$ and so $\Omega^{\ast}(\xi\mid
\nu)$ is also invariant under this transformation, leading to the scaling law
for the grand potential
\begin{equation}
\beta\Omega(\beta,q\mid\mu)=\gamma^{-2}\beta\Omega(\gamma^{-2}\beta
,\gamma^{1/2}q\mid\mu_{\gamma}) \label{3.21}%
\end{equation}
where $\mu_{\gamma}$ follows from (\ref{3.13})
\begin{equation}
\mu_{\gamma}=\gamma^{2}\mu(\gamma\mathbf{r}). \label{3.21a}%
\end{equation}
This last result may seem a bit surprising, since one na{\"\i}vely might
expect the same scaling as for the density (\ref{3.17}). But that does not
follow because of the assumed invertibility of $n$ and $\mu(\mathbf{r})$.
Instead (\ref{3.21a}) comes from
\begin{equation}
\nu(\mathbf{r}^{\ast})=\beta\mu(\mathbf{r}),\hspace{0.2in}\nu(\mathbf{r}%
^{\ast},\xi\mid n^{\ast})\rightarrow\nu(\mathbf{r}^{\ast\ast},\xi\mid n^{\ast
})=\nu(\frac{\lambda}{\lambda_{\gamma}}\mathbf{r}^{\ast},\xi\mid n^{\ast}%
)=\nu(\gamma\mathbf{r}^{\ast},\xi\mid n^{\ast}) \label{3.21b}%
\end{equation}
whence
\begin{equation}
\nu(\gamma\mathbf{r}^{\ast},\xi\mid n^{\ast})=\beta\mu(\mathbf{r}\gamma
)=\beta_{\gamma}\mu_{\gamma}(\mathbf{r}) \; . \label{3.21c}%
\end{equation}
Here the abbreviated notation is $\lambda_{\gamma}:= \lambda(\gamma^{-2}\beta)
= \gamma^{-1}\lambda(\beta)$ and $\beta_{\gamma}= \gamma^{-2}\beta$. The
result is (\ref{3.21a}).

Finally, the dimensionless internal energy and entropy of (\ref{3.15}) and
(\ref{3.16}) are also invariant under this transformation, giving their
scaling laws%
\begin{equation}
\beta U(\beta,q\mid n)=\beta\gamma^{-2}U(\gamma^{-2}\beta,\gamma^{1/2}q\mid
n_{\gamma}) \label{3.23}%
\end{equation}%
\begin{equation}
S(\beta,q\mid n)=S(\gamma^{-2}\beta,\gamma^{1/2}q\mid n_{\gamma}) \label{3.24}%
\end{equation}
For $q=0$ the results of reference \cite{DuftyTrickey11} for the
non-interacting case are recovered. An independent proof of these scaling laws
is given in Appendix \ref{apA} by direct coordinate scaling of the Hamiltonian
in (\ref{2.1}).

\subsection{Scaling properties of $\Omega(\beta\mid\mu,n)$ and $F(\beta\mid
\mu,n)$}

Now consider the dimensionless functionals of DFT. From (\ref{2.9}), the
dimensionless grand potential functional is
\begin{equation}
\Omega^{\ast}(\xi\mid\nu,n^{\ast})=\beta\Omega(\beta,q\mid\mu,n)=F^{\ast}%
(\xi\mid n^{\ast})-\int d\mathbf{r}^{\ast}\nu\left(  \mathbf{r}^{\ast}\right)
n^{\ast}(\mathbf{r}^{\ast}), \label{3.25}%
\end{equation}
as a functional of $n^{\ast}(\mathbf{r}^{\ast})$ for given $\nu\left(
\mathbf{r}^{\ast}\right)  $. Concurrently, consider the dimensionless free
energy density functional from (\ref{2.13}),
\begin{equation}
F^{\ast}(\xi\mid n^{\ast},\nu)=\beta F(\beta,q\mid n,\mu)=\Omega^{\ast}%
(\xi\mid\nu)+\int d\mathbf{r}^{\ast}\nu\left(  \mathbf{r}^{\ast}\right)
n^{\ast}(\mathbf{r}^{\ast}), \label{3.26}%
\end{equation}
as a functional of $\nu\left(  \mathbf{r}^{\ast}\right)  $ for given $n^{\ast
}(\mathbf{r}^{\ast})$. The dimensionless density $n^{\ast}\left(  {\mathbf{r}%
}^{\ast}\right)  $ and independent dimensionless local chemical potential
$\nu\left(  {\mathbf{r}}^{\ast}\right)  $ were defined at (\ref{3.8}) as
\begin{equation}
n^{\ast}(\mathbf{r}^{\ast})=\lambda^{3}n({\mathbf{r}}),\hspace{0.2in}%
\nu\left(  \mathbf{r}^{\ast}\right)  =\beta\mu\left(  \mathbf{r}\right)  \;.
\label{3.27}%
\end{equation}

The first terms on the right sides of (\ref{3.25}) and (\ref{3.26}) have the
invariance defined in (\ref{3.22}) and (\ref{3.21}), respectively. The
remaining term transforms as%
\[
\int d\mathbf{r}^{\ast}\nu\left(  \mathbf{r}^{\ast}\right)  n^{\ast
}(\mathbf{r}^{\ast})=\int d\mathbf{r}\beta\mu\left(  \mathbf{r}\right)
n({\mathbf{r}})
\]%
\begin{align}
&  \rightarrow\int d\mathbf{r}\beta_{\gamma}\mu_{\gamma}\left(  \mathbf{r}%
\right)  n_{\gamma}({\mathbf{r}})=\int d\mathbf{r}\left(  \mathbb{\gamma}%
^{-2}\beta\right)  \mathbb{\gamma}^{2}\mu\left(  \gamma\mathbf{r}\right)
\gamma^{3}n(\gamma{\mathbf{r}})\nonumber\\
&  =\int d\mathbf{r}\beta\mu\left(  \mathbf{r}\right)  n({\mathbf{r}}).
\label{3.29}%
\end{align}
Consequently, the DFT functionals $\Omega^{\ast}(\xi\mid\nu,n^{\ast})$ and
$F^{\ast}(\xi\mid n^{\ast},\nu)$ have the same invariance as the thermodynamic
functionals $\Omega^{\ast}(\xi\mid\nu)$ and $F^{\ast}(\xi\mid n^{\ast})$,%
\begin{equation}
\Omega(\beta,q\mid\mu,n)=\gamma^{-2}\Omega(\gamma^{-2}\beta,\gamma^{1/2}%
q\mid\mu_{\gamma},n_{\gamma}) \label{4.11}%
\end{equation}%
\begin{equation}
F(\beta,q\mid n,\mu)=\gamma^{-2}F(\gamma^{-2}\beta,\gamma^{1/2}q\mid
n_{\gamma},\mu_{\gamma}). \label{4.12}%
\end{equation}

\subsection{Exchange - correlation scaling}

The free energy functional $\beta F(\beta,q\mid n)$ typically is decomposed
into separate contributions for practical analysis%
\begin{equation}
F(\beta,q\mid n)=F^{(0)}(\beta\mid n)+E_{H}(q\mid n)+F^{xc}(\beta,q\mid n).
\label{3.32}%
\end{equation}
The first term on the right side $F^{(0)}$ is the free energy functional for
the non-interacting system,
\begin{equation}
F^{(0)}(\beta\mid n)=F(\beta,q=0\mid n), \label{3.33}%
\end{equation}
while the second is the Hartree energy%
\begin{equation}
E_{H}=q^{2}\int d\mathbf{r}d\mathbf{r}^{\prime}\frac{1}{|\mathbf{r}%
-\mathbf{r}^{\prime}|}\bigskip n(\mathbf{r})n(\mathbf{r}). \label{3.34}%
\end{equation}
The last term $F^{xc}$ is the exchange - correlation free energy, defined as
the residual free energy incorporating all exchange and correlation effects
due to interactions. It is easily seen that $\beta F^{(0)}(\beta\mid n)$ and
$\beta E_{H}(q\mid n)$ are invariant under the transformation defined by
(\ref{3.17}) and (\ref{3.19a}). Consequently, $\beta F^{xc}$ is invariant as
well%
\begin{equation}
\beta F^{xc}(\beta,q\mid n)=\gamma^{-2}\beta F^{xc}(\gamma^{-2}\beta
,\gamma^{1/2}q\mid n_{\gamma}). \label{3.35}%
\end{equation}

\section{Discussion}

\label{sec4}

At first reading, familiarity with ground-state DFT scaling laws may make
Eqs.\ (\ref{4.11}), (\ref{4.12}) a bit surprising because they both scale the
same way. Thus,
\begin{equation}
\beta Q(\beta,q\mid\mu,n)=\gamma^{-2}\beta Q(\gamma^{-2}\beta,\gamma
^{1/2}q\mid n_{\gamma},\mu_{\gamma}).\label{5.1}%
\end{equation}
for $Q$ any of $\Omega$, $F$, $U$, or ${\mathrm{T}}S$. Though not the more
commonly discussed ground-state DFT pattern (which involves inequalities),
such scaling has been discussed for the ground state by Levy; see Sec.\ 2.11
of Ref.\ \onlinecite{Levy87}. The relevant point there and here is the
inclusion of charge scaling, (\ref{3.19a}). Pittalis \textit{et al.}%
\cite{Pittalis11} arrived at the result (\ref{5.1}) by a route analogous to
Levy's ground-state procedure, namely by considering the coupling constant
scaling in the adiabatic connection procedure. This is the same charge scaling
as in (\ref{3.19a}) above but introduced separately from the uniform
coordinate scaling, (\ref{3.17}). The present analysis provides the
alternative perspective of invariance properties of the thermodynamical
functionals, (\ref{3.22}), (\ref{3.21}), (\ref{3.23}), and (\ref{3.24})
leading to the scaling of the DFT functionals. There is a similar parallel
with the recent discussion by Pribram-Jones and Burke of connecting charge and
temperature scaling \cite{PribramJonesBurke2015}.

Without the charge scaling but with temperature scaling as well as coordinate
scaling, the homogeneous scaling of the internal energy as $\gamma^{-2}$
(\ref{3.23}) does not hold, since the Coulombic contributions with unscaled
charge go as $\gamma^{-1}$, whereas the non-interacting parts $T_{s}$ (KE) and
${\mathrm{T}}S_{s}$ (entropy) scale as $\gamma^{-2}$ (obviously irrespective
of charge scaling) \cite{DuftyTrickey11}. Since the exchange contribution
$U_{x}(\beta,q\mid\mu,n)$ to the internal energy functional is defined in DFT
(both finite-T and ground-state) as a Coulomb energy (from the 1-body reduced
density matrix), its scaling is the same as the Hartree contribution
$U_{H}(\beta,q-\mu,n)$ : both go as $\gamma^{-2}$ with charge scaling,
$\gamma^{-1}$ without. As with the ground-state theory, the difference between
the scaling procedures shows up in the correlation internal energy
$U_{c}(\beta,q\mid\mu,n)$ because it has both a Coulombic contribution
\begin{equation}
U^{c}_{Coul}(\beta,q\mid\mu,n)=W_{ee}(\beta,q\mid\mu,n)-U_{H}(\beta,q\mid
\mu,n)-U_{x}(\beta,q\mid\mu,n) \label{UcCouldefn}%
\end{equation}
(with $W_{ee}$ the entire Coulomb internal energy functional) and two
contributions that scale as does a kinetic energy, namely,
\begin{align}
{\mathcal{T}}^{c}(\beta,q\mid\mu,n)  &  ={\mathcal{T}}(\beta,q\mid
\mu,n)-{\mathcal{T}}_{s}(\beta,q\mid\mu,n)\label{Tcdefn}\\
{\mathrm{T}}{\mathcal{S}}^{c}(\beta,q\mid\mu,n)  &  ={\mathrm{T}}{\mathcal{S}%
}(\beta,q\mid\mu,n)-{\mathrm{T}}{\mathcal{S}}_{s}(\beta,q\mid\mu,n)\;.
\label{Scdefn}%
\end{align}
Here $\mathcal{T}$, $\mathcal{S}$ are the kinetic energy and entropy
functionals and subscript $s$ labels their non-interacting (Kohn-Sham)
counterparts. Thus,
\begin{equation}
U^{c}(\beta,q\mid\mu,n)=U^{c}_{Coul}(\beta,q\mid\mu,n)+{\mathcal{T}}^{c}%
(\beta,q\mid\mu,n)+{\mathrm{T}}{\mathcal{S}} ^{c}(\beta,q\mid\mu,n)\;.
\label{UcDefn}%
\end{equation}
Without charge scaling, this combination leads \cite{Pittalis11} to
inequalities such as
\begin{equation}
\Omega^{c}(\beta,q\mid n_{\gamma},\mu_{\gamma})\geq\gamma\Omega^{c}%
(\gamma^{-2}\beta,q\mid n,\mu)\;,\;\gamma\geq1 \label{PittalisEq28}%
\end{equation}
for the correlation contribution to the grand potential. In contrast, with
charge scaling, the equality (\ref{5.1}) holds. Both forms are potentially
useful for constraining the design of approximate functionals.

\section{Acknowledgment}

\label{sec6} This work was supported under U.S DOE Grant DE-SC0002139. We
thank Stefano Pitallis and Kieron Burke for useful discussions.

\bigskip

\appendix

\section{Scaling of $\Omega(\beta,q\mid\mu)$ and $F(\beta,q\mid n)$}

\label{apA}The scaling properties for the functionals of thermodynamics follow
directly from the statistical mechanical definition (\ref{2.1}). In our
earlier work on the non-interacting functionals \cite{DuftyTrickey11} a
unitary operator was constructed that implements scaling of the particle
coordinate and momentum operators%
\begin{equation}
\widehat{U}_{\gamma}A(\widehat{\mathbf{q}}_{\alpha}\mathbf{,}\widehat
{\mathbf{p}}_{\alpha})\widehat{U}_{\gamma}^{-1}=A(\gamma\widehat{\mathbf{q}%
}_{\alpha}\mathbf{,}\gamma^{-1}\widehat{\mathbf{p}}_{\alpha}),\hspace
{0.25in}0<\gamma<\infty\,, \label{a.1}%
\end{equation}
where $\gamma$ again is a positive, real-valued scalar. The detailed form of
$\widehat{U}_{\gamma}$ is not required here; the interested reader is referred
to Appendix B of reference \cite{DuftyTrickey11}. Then from the cyclic
invariance of the trace Eq.\ (\ref{2.1}) of the main text can be written
\begin{align}
\beta\Omega(\beta,q  &  \mid\mu)=-\ln\sum_{N=0}^{\infty}Tr^{(N)}\widehat
{U}_{\gamma}e^{-\beta\left(  \widehat{H}-\int d\mathbf{r}\mu(\mathbf{r}%
)\widehat{n}(\mathbf{r})\right)  }\,\widehat{U}_{\gamma}^{-1}\nonumber\\
&  =-\ln\sum_{N=0}^{\infty}Tr^{(N)}e^{-\widehat{U}_{\gamma}\beta\left(
\widehat{H}-\int d\mathbf{r}\mu(\mathbf{r})\widehat{n}(\mathbf{r})\right)
\widehat{U}_{\gamma}^{-1}}\,, \label{a.2}%
\end{align}
and by application of (\ref{a.1})
\begin{align}
\widehat{U}_{\gamma}\beta\left(  \widehat{H}-\int d\mathbf{r}\mu
(\mathbf{r})\widehat{n}(\mathbf{r})\right)  \widehat{U}_{\xi}^{-1}  &
=\beta\gamma^{-2}\left(  \sum_{\alpha=1}^{N}\left(  \frac{\widehat{p}_{\alpha
}^{2}}{2m}-\gamma^{2}\mu\left(  \gamma\hat{\mathbf{q}}_{\alpha}\right)
\right)  +\tfrac{1}{2}\gamma q^{2}\sum_{\alpha\neq\eta=1}^{N}\frac{1}%
{|\hat{\mathbf{q}}_{\alpha}-\hat{\mathbf{q}}_{\eta}|}\right) \nonumber\\
&  =\beta_{\gamma}\left(  \sum_{\alpha=1}^{N}\frac{\widehat{p}_{\alpha}^{2}%
}{2m}+\tfrac{1}{2}q_{\gamma}^{2}\sum_{\alpha\neq\eta=1}^{N}\frac{1}%
{|\hat{\mathbf{q}}_{\alpha}-\hat{\mathbf{q}}_{\eta}|}-\int d\mathbf{r}%
\mu_{\gamma}(\mathbf{r})\widehat{n}(\mathbf{r})\right)  \label{a.3}%
\end{align}
with
\begin{equation}
\beta_{\gamma}=\beta\gamma^{-2},\hspace{0.2in}q_{\gamma}=\gamma^{1/2}%
q,\hspace{0.2in}\mu_{\gamma}(\mathbf{r})=\gamma^{2}\mu\left(  \gamma
\mathbf{r}\right)  \label{a.4}%
\end{equation}
Consequently, the desired scaling property is obtained%
\begin{equation}
\beta\Omega(\beta,q\mid\mu)=\beta_{\gamma}\Omega(\beta_{\gamma},q_{\gamma}%
\mid\mu_{\gamma}). \label{a.5}%
\end{equation}

To interpret the transformed local chemical potential $\mu_{\gamma}$ note that
it gives rise to the density
\begin{equation}
n_{\gamma}\left(  \mathbf{r,}\beta_{\gamma},q_{\gamma}\mid\mu_{\gamma}\right)
=\gamma^{3}\frac{\delta\Omega(\beta_{\gamma},q_{\gamma}\mid\mu_{\gamma}%
)}{\delta\mu_{\gamma}\left(  \mathbf{r}\right)  }. \label{a.6}%
\end{equation}
This density is related to that in the original variables by%
\begin{equation}
e^{\beta_{\gamma}\Omega(\beta_{\gamma},q_{\gamma}\mid\mu_{\gamma})}\sum
_{N=0}^{\infty}Tr^{(N)}\widehat{n}(\mathbf{r})e^{-\widehat{U}_{\gamma}%
\beta\left(  \widehat{H}-\int d\mathbf{r}\mu(\mathbf{r})\widehat{n}%
(\mathbf{r})\right)  \widehat{U}_{\gamma}^{-1}}=e^{\beta\Omega(\beta,q\mid
\mu)}\sum_{N=0}^{\infty}Tr^{(N)}\widehat{U}_{\xi}^{-1}\widehat{n}%
(\mathbf{r})\widehat{U}_{\xi}e^{-\beta\left(  \widehat{H}-\int d\mathbf{r}%
\mu(\mathbf{r})\widehat{n}(\mathbf{r})\right)  }, \label{a.7}%
\end{equation}
or%
\begin{equation}
n\left(  \mathbf{r,}\beta_{\gamma},Q_{\gamma}\mid\mu_{\gamma}\right)
=\gamma^{3}n\left(  \gamma\mathbf{r,}\beta,q\mid\mu\right)  , \label{a.8}%
\end{equation}
i.e., it is the transformed density $n_{\gamma}(\mathbf{r})$.

Now consider the Legendre transform for the new free energy $F(\beta_{\gamma
},Q_{\gamma}\mid n_{\gamma})$%
\begin{align}
\beta_{\gamma}F(\beta_{\gamma},q_{\gamma}  &  \mid n_{\gamma})=\beta_{\gamma
}\Omega(\beta_{\gamma},q_{\gamma}\mid\mu_{\gamma})+\beta_{\gamma}\int
d\mathbf{r}\mu_{\gamma}\left(  \mathbf{r}\right)  n_{\gamma}\left(
\mathbf{r}\right) \nonumber\\
&  =\beta\Omega(\beta,Q\mid\mu)+\beta\int d\mathbf{r}\mu\left(  \xi
\mathbf{r}\right)  \xi^{3}n\left(  \xi\mathbf{r}\right) \nonumber\\
&  =\beta F(\beta,Q\mid n). \label{a.9}%
\end{align}
This is the expected invariance corresponding to that for $\beta\Omega
(\beta,q\mid\mu)$ in (\ref{a.5}).

\bigskip


\begin{thebibliography}{99}                                                                                               %


\bibitem {Savin96}A.\ Savin, in \textit{Recent Development and Application of
Modern Density Functional Theory}, Theoret.\ and Comput.\ Chem.\ vol.\ 4,
J.M.\ Seminario ed. (Elsevier, Amsterdam, 1996) p.\ 327.

\bibitem {Mermin65}N.D.~Mermin, Phys.\ Rev.\ \textbf{137}, A1441 (1965).

\bibitem {YangParrBook}R.G.\ Parr and W.\ Yang, \textit{Density Functional
Theory of Atoms and Molecules} (Oxford, New York, 1989).

\bibitem {DreizlerGrossBook}R.M.\ Dreizler and E.K.U.\ Gross, \textit{Density
Functional Theory} (Springer-Verlag, Berlin, 1990).

\bibitem {KryachkoLudena}E.S.~Kryachko and E.V.~Lude\~na, \textit{Energy
Density Functional Theory of Many-Electron Systems} (Kluwer, Dordrecht, 1990).

\bibitem {Eschrig}H.~Eschrig, \textit{The Fundamentals of Density Functional
Theory} (Teubner, Stuttgart, 1996).

\bibitem {LudenaKarasiev2002}E.V.\ Lude\~na and V.V.\ Karasiev, in
\textit{Reviews of Modern Quantum Chemistry: a Celebration of the
Contributions of Robert Parr}, K.D.\ Sen ed.\ (World Scientific, Singapore,
2002) 612.

\bibitem {FilohaisEtAl03}\emph{A Primer in Density Functional Theory},
C.~Fiolhais, F.~Nogueira, and M.A.L.~Marques eds.\ (Springer, Berlin, 2003).

\bibitem {EngelDreizler11}\emph{Density Functional Theory: An Advanced Course}
E.\ Engel and R.M.\ Dreizler (Springer, Heidelberg, 2011).

\bibitem {Evans92}R. Evans, in \emph{Fundamentals of Inhomogeneous Fluids},
edited by D. Henderson (Marcel Dekker, NY, NY 1992).

\bibitem {Lowen94}H.\ L\"{o}wen, Phys.\ Rep.\ \textbf{237}, 249 (1994).

\bibitem {Dharma95}C.\ Dharma-wardana and F.\ Perrot in \emph{Density
Functional Theory}, E.K.U.\ Gross and R.M.\ Dreizler eds.\ (Plenum Press, NY,
1995) 625.

\bibitem {Lutsko10}J.\ Lutsko, in Adv.\ Chem.\ Phys.\ \textbf{144}, 1 (2010).

\bibitem {WDM}\emph{Basic Research Needs in High Energy Density Laboratory
Physics}, R.\ Rozner, D.\ Hammer, and T.\ Rothman (U.S.\ Dept.\ of Energy,
2010), Chapter 6 and references therein.

\bibitem {IPAMbook}\emph{Frontiers and Challenges in Warm Dense Matter}, F.
Graziani, M.P. Desjarlais, R. Redmer, and S.B. Trickey eds., (Springer,
Heidelberg, 2014).

\bibitem {LevyPerdew85}M.~Levy and J.P.~Perdew, Phys.\ Rev.\ A \textbf{32},
2010 (1985).

\bibitem {DuftyTrickey11}J.W.\ Dufty and S.B.\ Trickey, Phys.\ Rev.\ B
\textbf{84}, 125118 (2011).

\bibitem {Pittalis11}S.\ Pittalis, C.R.\ Proetto, A.\ Floris, A.\ Sanna,
C.\ Bersier, K.\ Burke, and E.K.U.\ Gross, Phys.\ Rev.\ Lett.\ \textbf{107},
163001 (2011).

\bibitem {Thermo}The relationship of DFT to thermodynamics and Legendre
transforms is discussed in E.H.\ Lieb, Int.\ J.\ Quantum Chem. \textbf{24},
243 (1983); N.\ Argaman and G.\ Markov, Am.\ J.\ Phys.\ \textbf{68}, 69
(2000); J-M.\ Caillol, J.\ Phys.\ A: Math.\ Gen.\ \textbf{35}, 4189 (2002);
J.\ Chihara and M.\ Yamagiwa, Prog.\ Theor.\ Phys.\ \textbf{111}, 339 (2004);
R.\ Balawender and A.\ Holas, ``The thermodynamic extension of density
functional theory: I'', preprint 2009, arXiv:0901.1060; and in Ref.\ \onlinecite{Eschrig10}.

\bibitem {Eschrig10}H.\ Eschrig, Phys.\ Rev.\ B \textbf{82}, 205120 (2010).

\bibitem {Caillol02}See Caillol in \cite{Thermo}.

\bibitem {Ruelle1969}\textit{Statistical Physics: Rigorous Results},
D.\ Ruelle (Benjamin, NY, 1969), proposition 2.5.5. The functional at issue
here differs from the thermodynamic functional only by a linear functional,
hence does not change concavity.

\bibitem {H_Kohn}P.  Hohenberg and W. Kohn, Phys. Rev. \textbf{136}, B864--871 (1964).

\bibitem {Liebref}Also see Lieb in \cite{Thermo}.

\bibitem {ChakrabortyEtAl2015}D.\ Chakraborty, J.W.\ Dufty, and
V.V.\ Karasiev, Adv.\ Quantum Chem.\ \textbf{71}, 11 (2015).

\bibitem {Levy87}M.~Levy in \emph{Single-Particle Density in Physics and
Chemistry}, N.H.~March ed. (Academic, London and San Diego, 1987) 45.

\bibitem {PribramJonesBurke2015}\textquotedblleft Connection Formula for
Thermal Density Functional Theory\textquotedblright, A. Pribram-Jones and K.
Burke, arXiv:1509.03060.
\end{thebibliography}
\end{document}